\documentclass[aps, pra, showpacs,amsfonts,twocolumn, amsmath, amssymb, superscriptaddress, floatfix]{revtex4}
\usepackage{graphicx}
\usepackage{dcolumn}
\usepackage{bm}
\begin{document}

\title{Anderson localization in optical lattices with speckle disorder}

\author{Serpil Sucu}\affiliation{Department of Physics, Trakya University, 22030 Edirne, Turkey}
\author{Saban Aktas}\affiliation{Department of Physics, Trakya University, 22030 Edirne, Turkey} 
\author{S. Erol Okan}\affiliation{Department of Physics, Trakya University, 22030 Edirne, Turkey} 
\author{Zehra Akdeniz}\affiliation{Piri Reis University, 34940 Tuzla-Istanbul, Turkey} 
\author{Patrizia Vignolo}\affiliation{Universit\'e de Nice - Sophia Antipolis, Institut non Lin\'eaire de Nice, CNRS, 1361 route des Lucioles, 06560 Valbonne, France}

\begin{abstract}
We study the localization properties of non-interacting waves
propagating in a speckle-like potential superposed on
a one-dimensional lattice. Using a 
decimation/renormalization procedure, we estimate the localization length
for a tight-binding Hamiltonian where site-energies are square-sinc-correlated
random variables. By decreasing the width of the correlation function, the 
disorder patterns approaches a $\delta$-correlated disorder, and the 
localization length becomes almost energy-independent in the 
strong disorder limit. We show that this regime can be reached for 
a size of the speckle grains of the order of (lower than) four lattice steps.
\end{abstract}


\maketitle

\section{Introduction}
The Anderson model was proposed first to explain the absence of electronic 
diffusion in certain random lattices \cite{Ande58} and later to explain 
the absence 
of diffusion of light in certain amorphous materials \cite{Ande85}.
In the first case, the electron energy is lower than the maxima of the 
lattice potential and the particles diffuse by tunnelling. In the second 
case the particles (photons in \cite{Ande85}) energy is higher than the potential and the 
particles are ``free''.
In both cases, in the absence of disorder, the eigenstates are delocalized
states (Bloch waves in the first regime, plane waves in the second).
The presence of a $\delta$-correlated random potential freezes the wave
propagation at a length of $\mathcal{L}_{loc}$, the localization length,
always in  one (1D) and two dimension (2D), and 
depending on the disorder strength and the energy wave also in three dimension (3D) \cite{Ande79}.
In the last few years the ultracold atom community
has devoted a large effort to the experimental realization
of Anderson localization.
Anderson localization in the tight-binding regime was observed 
in momentum space with kicked-rotor set-ups, in 1D \cite{Moore1995},
and 3D \cite{Chabe2008}, and in real space by using a quasi-periodic potential 
(thus not strictly-speaking a random potential).
The key-ingredient for the experimental study of Anderson localization 
of ultracold atoms in the ``free''-particle regime has been the 
speckle potential \cite{Bouyer2010}. 
By using this handable optical potential, Anderson
localization was observed in
1D \cite{Billy2008} and in 3D \cite{Kondov2011,Jendrzejewski2011}, and anomalous diffusion
was observed in 2D \cite{Robertde2010}. 
The auto-correlation function of the speckle potential is a square-sinc,
thus it decays algebraically as a long-range correlated disorder, but
is also characterized by a finite size correlation length 
$w$, corresponding to the width of central bump of the square-sinc function.

 In 1D, the presence of disorder patterns with auto-correlation 
functions decaying algebraically can mimic
the presence of a mobility edge \cite{Tessieri2002,Kuhl,Gurevich2009,Luga09} and 
can even enhance localization, as shown in a microwave experiment \cite{Kuhl}. 
Both phenomena are due to the fact
that the disorder spectrum is non-zero in a finite momentum interval.

Very recently Semmler and coworkers \cite{Semmler2010} have studied the 
phase diagram of correlated fermions in 2D and 3D optical lattices and 
in the presence of a speckle potential. From the analysis of the local 
Density Of States (DOS) they identify an Anderson-Mott and a Mott 
localized phase as functions of the interaction strength and 
the strength of the speckle potential.  
In this article we analyze the possibility of observing Anderson localization
of a non-interacting wave, for example a non-interacting 
Bose-Einstein condensate \cite{Roati2008}, in a speckle potential 
{\it superposed}
to a 1D lattice potential.
By using a decimation/renormalization scheme \cite{Farchioni1992a} we
analyze how the DOS of a lattice is modified by the presence of the speckle,
and we estimate the localization length $\mathcal{L}_{loc}$ as function of
the disorder strength and of the width $w$ of the auto-correlation function.
The speckle potential is introduced as an on-site disorder which has 
statistical properties which are the same as a genuine speckle potential. 
This is illustrated in Sec. \ref{sec:model}. In
Sec. \ref{sec:results} we remind the reader of
the decimation/renormalization procedure exploited to compute
the DOS and $\mathcal{L}_{loc}$. Our results 
show how the efficacy of the speckle potential to localize increases
by increasing the disorder strength and by decreasing the correlation 
length $w$. These results can be a guide to choosing the experimental parameters
to observe Anderson localization in the tight binding regime with speckle
disordered patterns.
 
\section{The model}
\label{sec:model}
To study the effect of a speckle potential in the presence of a 1D lattice 
on matter-wave transport
we use the 1D Tight-Binding (TB) Hamiltonian,
\begin{equation}
H=\sum_{i=1}^{n_s}  
E_i |\,i\rangle\langle i\,|+
\sum_{i=1}^{n_s-1}t(|\,i\rangle\langle i+1\,|+|\,i+1\rangle\langle i\,|)
\label{Hamiltonian}
\end{equation}
where $n_s$ is the number of sites, $E_i$ the energy at the site $i$. The hopping term $t$ is chosen site-independent.
The effect of the speckle potential is introduced in the on-site energy distribution by setting
\begin{equation}
\mathcal{C}_\ell=\langle \delta E_i \delta E_{i+\ell}\rangle=s^2\left(\frac{\sin(2\pi\ell/w)}{2\pi\ell/w}\right)^2,
\label{diffr}
\end{equation}
where $\delta E_i=E_i-\langle E_i\rangle$ is the fluctuation of $E_i$ with respect to the mean value $\langle E_i\rangle$, $s=\sqrt{\langle(\delta E_i)^2\rangle}$ is the disorder strength and $w$ the width, in the units of the lattice step $d$, of the correlation function in Eq. (\ref{diffr}).
The disorder spectrum is not uniform as in the Anderson model \cite{Ande58}, but
is described by the triangular function \cite{Billy2008}
\begin{equation}
\mathcal{S}_k\propto s^2(\kappa-|k|)\theta(\kappa-|k|)
\label{spectral}
\end{equation}
where $\kappa=4\pi/w$, and $\theta(x)$ is the Heaviside function.
\subsection{Generation of the disordered potential}
We use the Fourier Filtering Method (FFM) \cite{Peng1991,Prakash1992,Makse1996} to generate the
disorder pattern described by the correlation function (\ref{diffr}).
First we generate a sequence of $N$ $\delta$-correlated random numbers $\{u_j\}$, with
$j=1,\dots N$ from a uniform distribution centered in zero and of width 1.
The second step is the generation of the desired $\{E_j\}$ distribution by ``filtering'', in Fourier space, the uniform 
distribution $\{u_j\}$. The filter being the spectral function $\mathcal{S}_k$, the $E_j$'s are evaluated directly from 
the expression
\begin{equation}
E_j=\frac{1}{N_k}\sum_{j_k=0}^{N_k}
\sum_{m=1}^N\sqrt{\mathcal{S}_k}e^{ik(m-j)}u_m,
\end{equation}
where $N_k=8N/w$ et $k=-\kappa+(\pi/N) j_k$. By construction $\langle E_j\rangle=0$, namely $\delta E_j=E_j$, and $\langle E_jE_{j+\ell}\rangle$ verifies Eq. (\ref{diffr}) in the limit $N\rightarrow\infty$.
A different choice of $\langle E_j\rangle$ would just shift the 
zero of the energy. 
\section{Numerical results: the DOS and the localization length}
\label{sec:results}
In the continuous limit, the single-particle DOS for an optical speckle
has been studied in \cite{Falco2010}. 
In the presence of a lattice (and in the absence of the speckle potential), the 
low-energy single-particle DOS has a typical saddle shape with two horns
that correspond respectively to the center and the edge of 
the first Brillouin zone.  
To evaluate how the speckle potential modifies the 
DOS of the lattice, we compute the DOS, $\mathcal{N}(E)$, 
regarding the Hamiltonian (\ref{Hamiltonian})
by using the Kirkman-Pendry relation \cite{Kirkman1984}
\begin{equation}
\mathcal{N}(E)=\lim_{\varepsilon\rightarrow 0^+}\frac{1}{\pi}{\rm Im}\left\{
\frac{\partial 
\ln [G_{1,n_s}(E+i\varepsilon)]}{\partial E}\right\}.
\end{equation}
Here $G(E)=(E-H)^{-1}$ is the Green's function related to the Hamiltonian
$H$ at energy $E$, and $G_{i,j}(E)=\langle i|G(E)|j \rangle$. 
With the aim of computing the matrix element $G_{1,n_s}(E)$,
we reduce the dimensionality of the system by evaluating the effective
Hamiltonian 
\begin{equation}
\tilde{H}=\tilde{E}_1 |\,1\rangle\langle 1\,|+\tilde{E}_{n_s} 
|\,n_s\rangle\langle n_s\,|+
\tilde{t}(|\,1\rangle\langle n_s+1\,|+|\,n_s+1\rangle\langle 1\,|),
\label{H2}
\end{equation}
where $\tilde{E}_1$, $\tilde{E}_{n_s}$ and $\tilde t$
are functions of the energy $E$ and of the Hamiltonian elements
of the decimated states (2, 3, \dots $n_s-1$) \cite{Farchioni1992a,Vign03}.
The Green's function of the effective Hamiltonian (\ref{H2}), 
$\tilde{G}(E)=(E-\tilde{H})^{-1}$, coincides with $G(E)$ in the subspace
$\{1,n_s\}$ by construction.

\begin{figure}
\includegraphics[width=0.49\linewidth,clip=true]{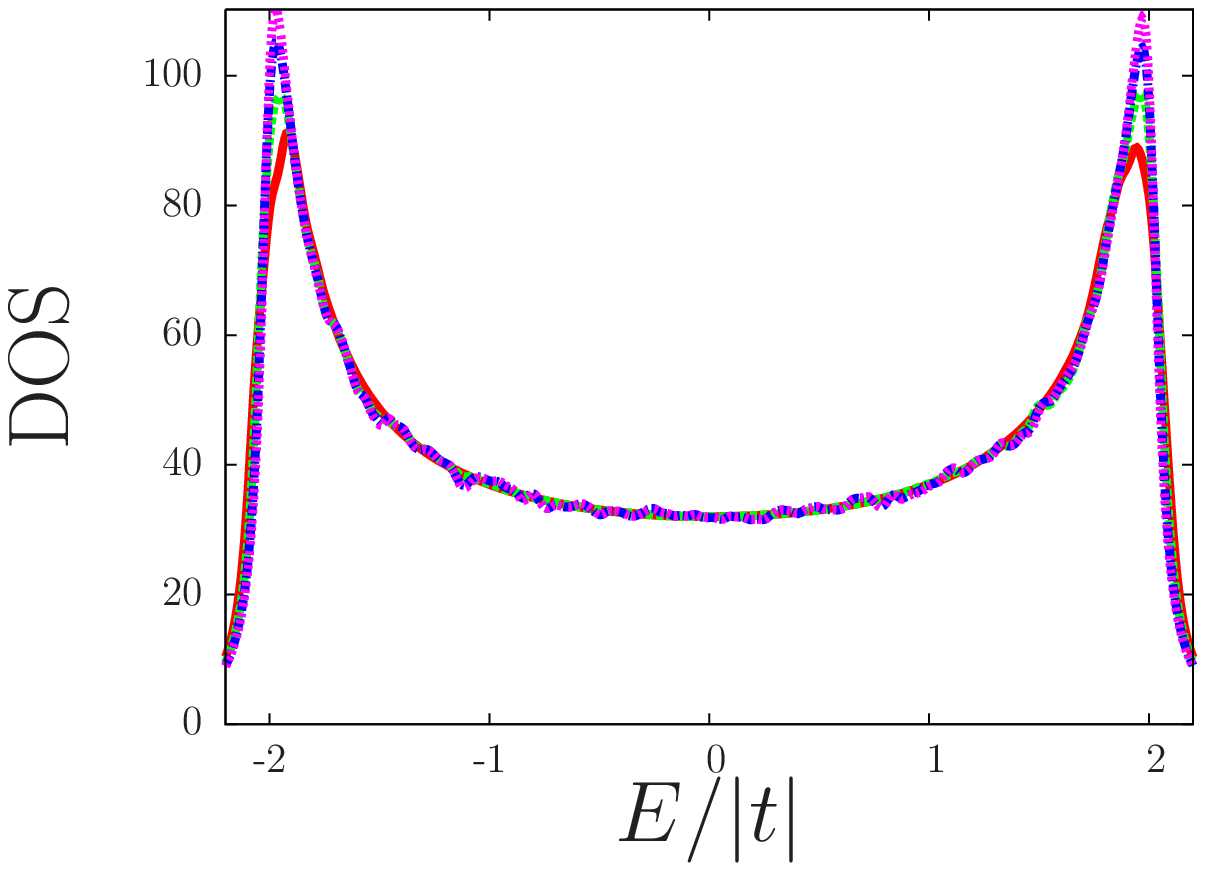}
\includegraphics[width=0.49\linewidth,clip=true]{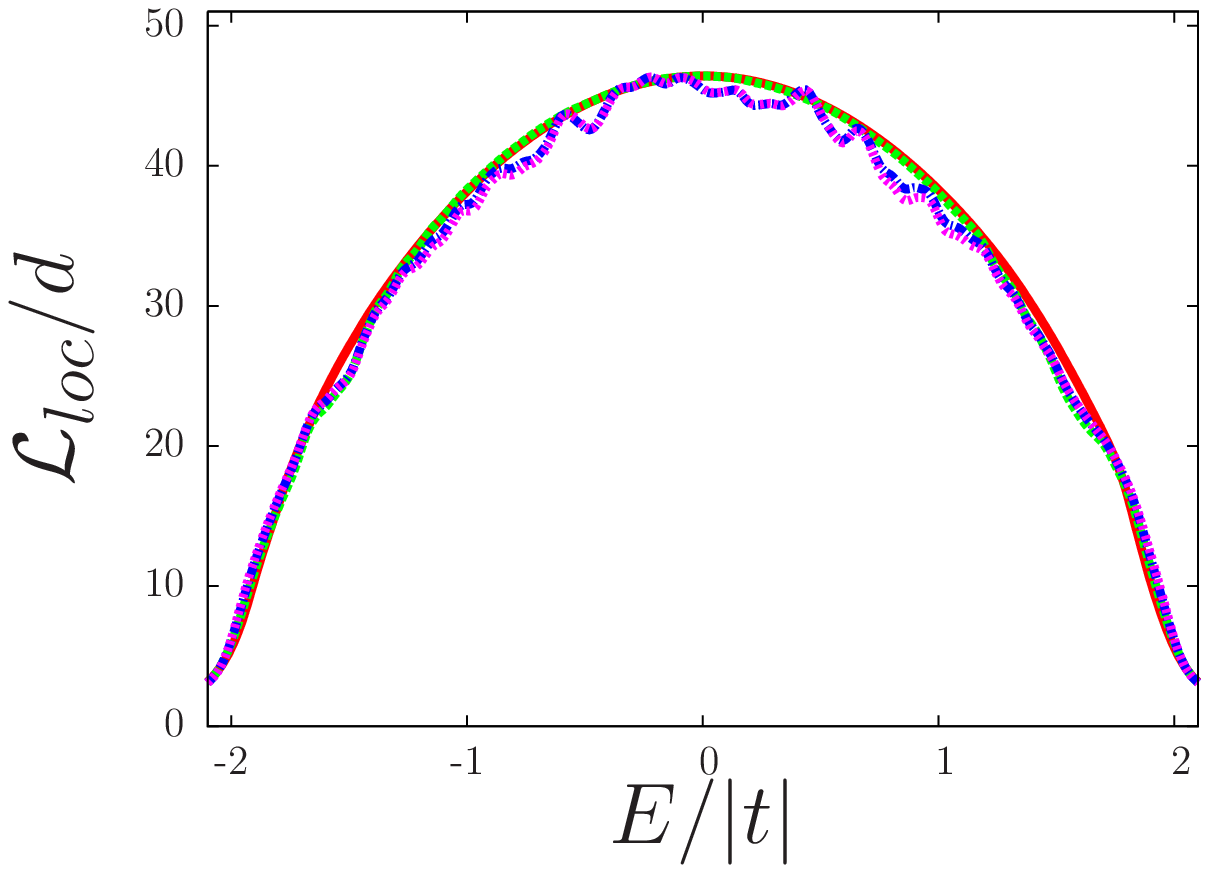}\\
\includegraphics[width=0.49\linewidth,clip=true]{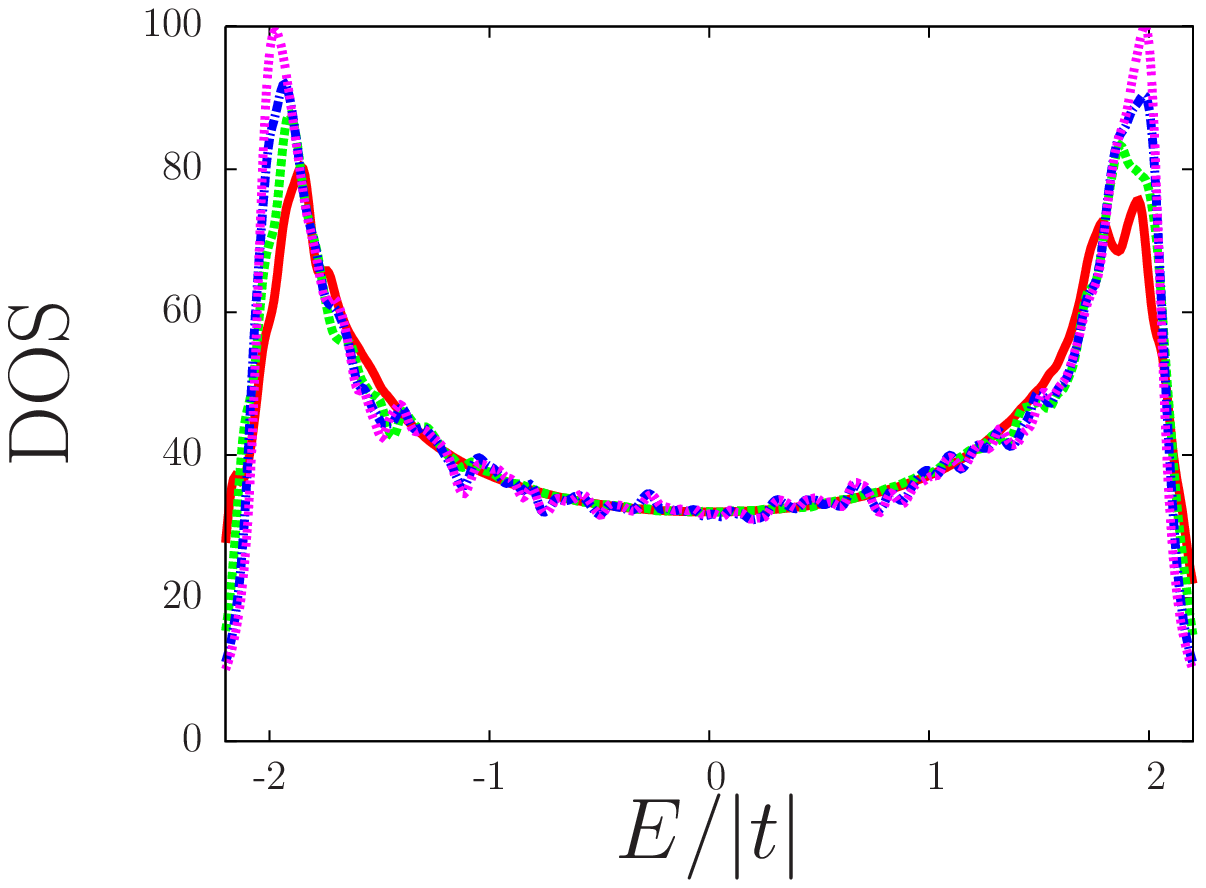}
\includegraphics[width=0.49\linewidth,clip=true]{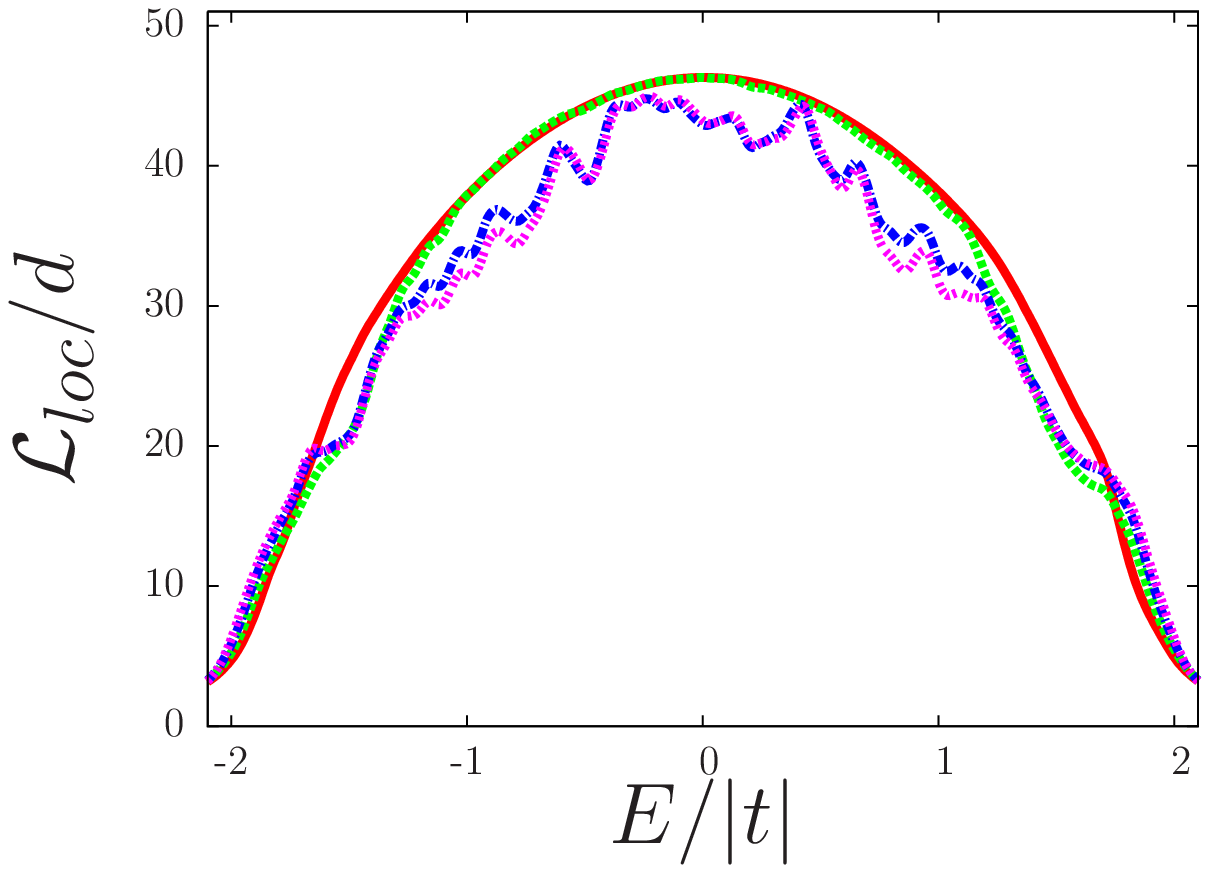}\\
\includegraphics[width=0.49\linewidth,clip=true]{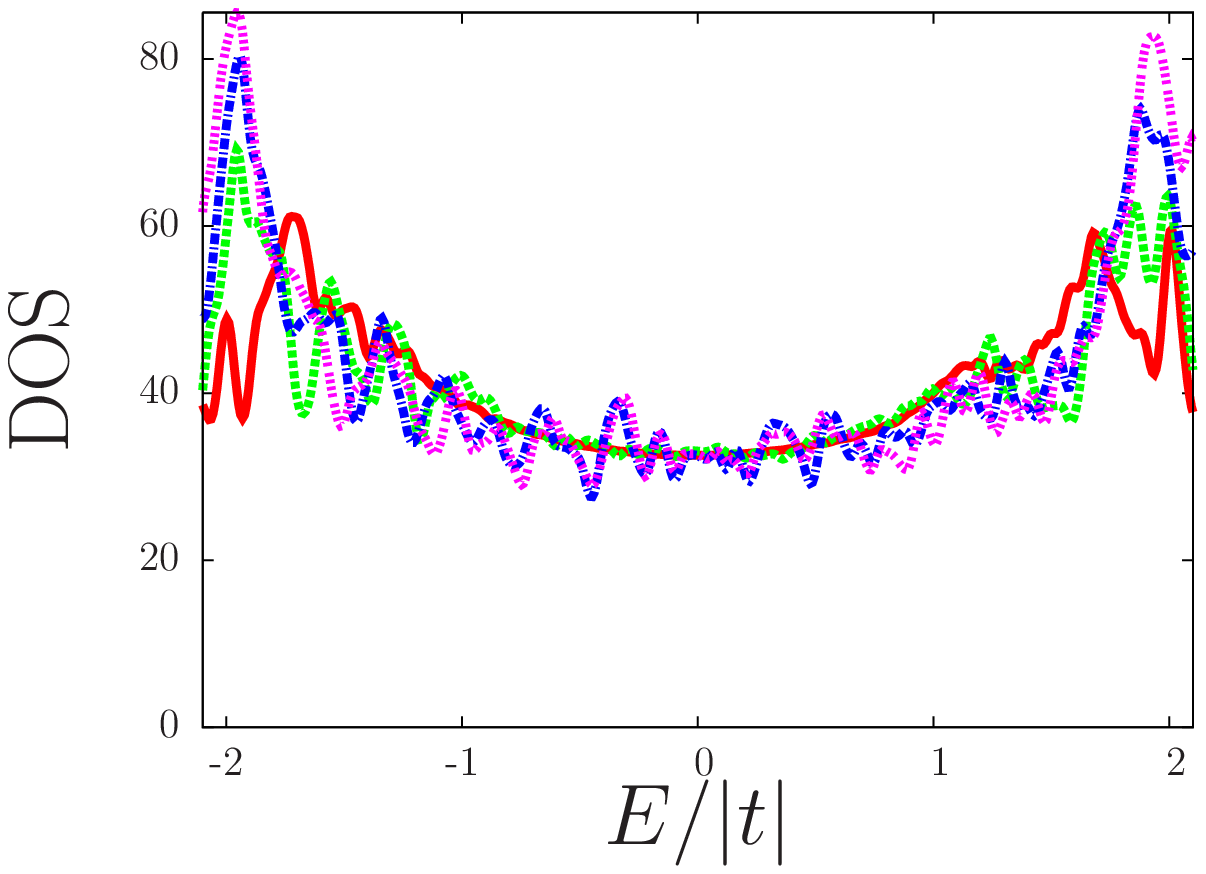}
\includegraphics[width=0.49\linewidth,clip=true]{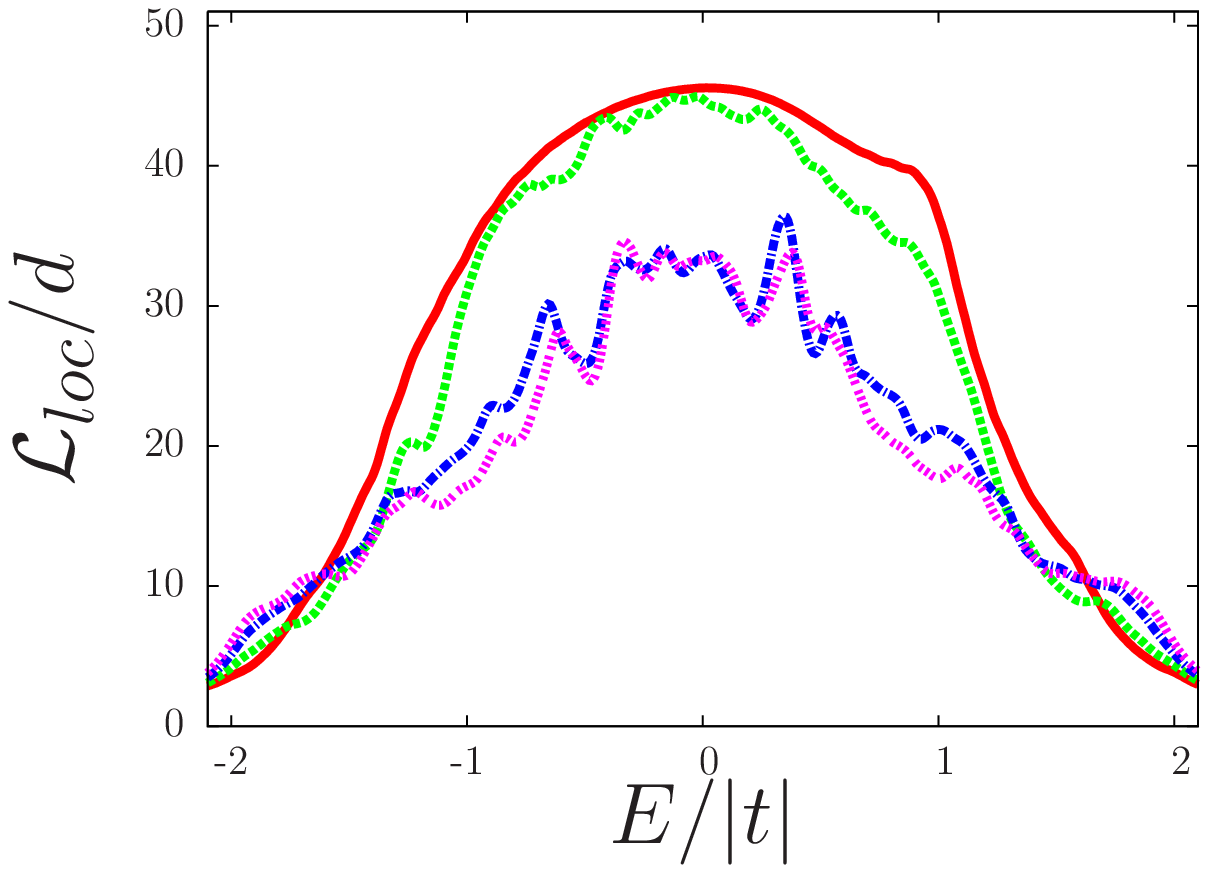}\\
\includegraphics[width=0.49\linewidth,clip=true]{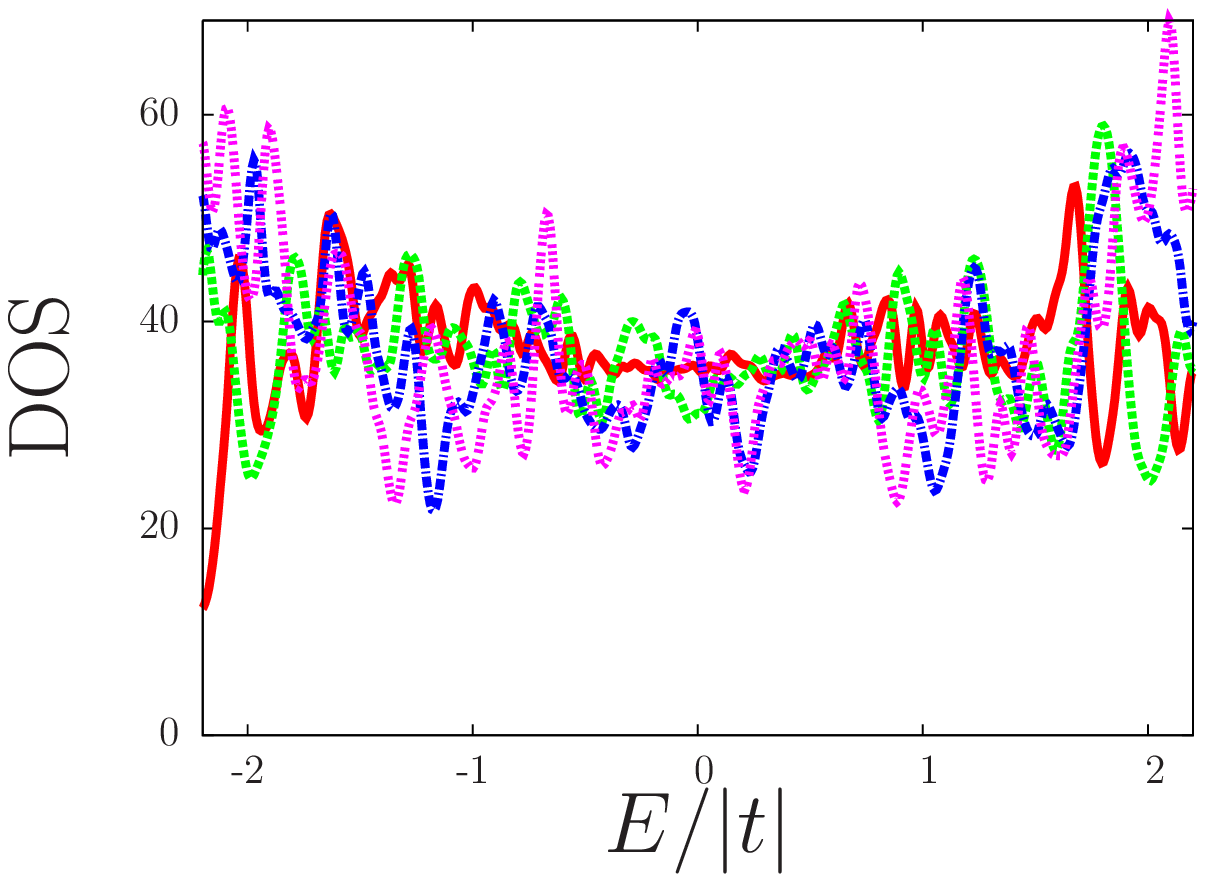}
\includegraphics[width=0.49\linewidth,clip=true]{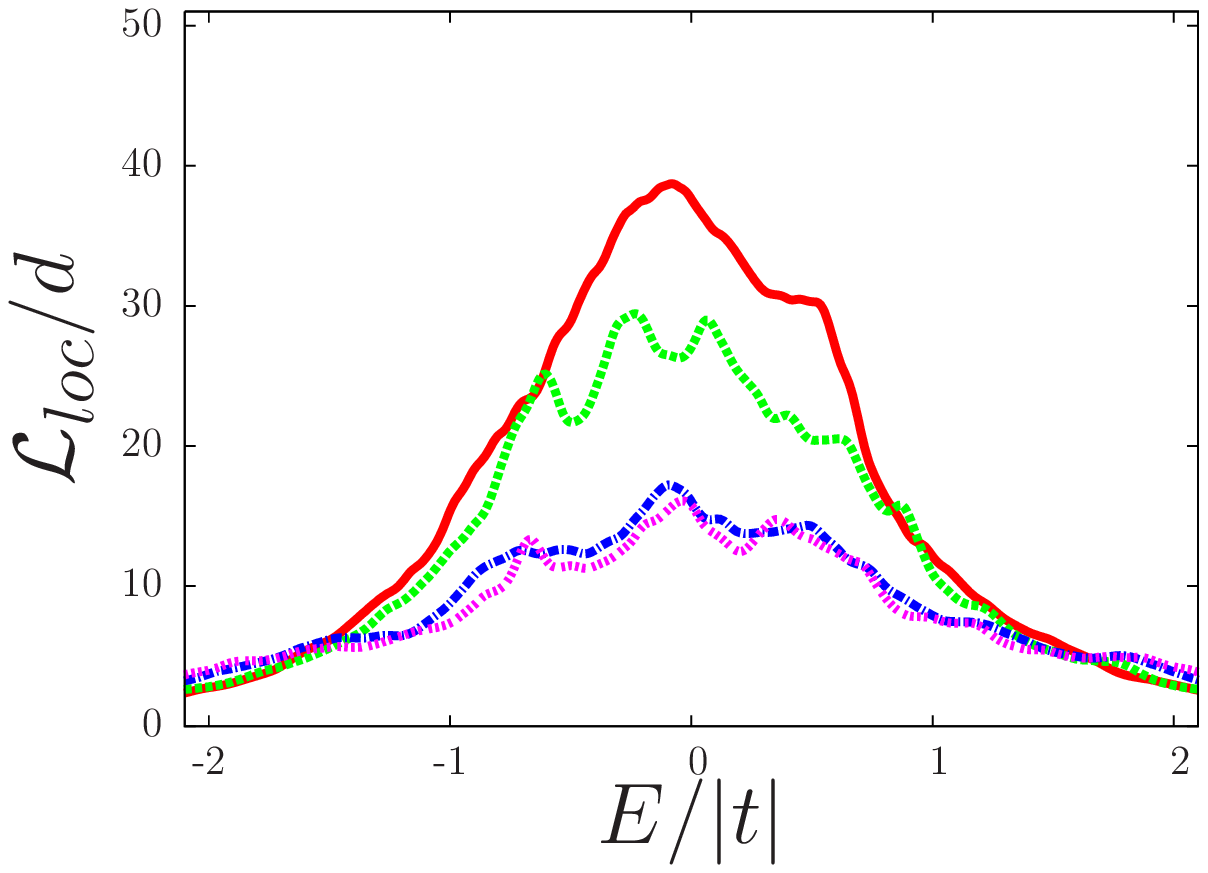}
\caption{\label{fig1} (Color online) 
DOS and localization length $\mathcal{L}_{loc}$ 
(in units of the lattice step $d$) as functions of the particle energy 
(in units of the energy hopping) for a lattice of 200 sites where site 
energies are $\mathcal{C}_\ell$-distributed random variables 
[see Eq. (\ref{diffr})]. Each curve correponds to an overage over 100 
configurations.
The different lines correspond to: $w=4\pi$ (continuous red line), $2\pi$ (dashed green line), $\pi$ (point-dashed blue line) and $2\pi/3$ (pointed magenta line).
From top to bottom: $s/|t|=1, 2,5$ and 10.}
\end{figure}

The numerical results for the DOS
as a function of the energy in units of $|t|$ are shown in the first column
of Fig. \ref{fig1}. One can observe that for large values of $w$, the speckle
disorder mainly affects the edge states of the DOS of the
underlying perfect chain, while
for smaller values of $w$ the disorder mainly influences the central 
part of the
spectrum.

The presence of the disorder modifies not only the DOS, but also the nature
of the states, from extended to localized.
In the continuous limit, the presence of the correlations 
described by Eq. (\ref{diffr}) does not
destroy localization but deeply modifies
the behaviour of the localization length as a function 
of the energy \cite{Tessieri2002,Gurevich2009,Luga09}.
To study the behaviour of the localization length $\mathcal{L}_{loc}(E)$
in the tight-binding regime, we compute
the Lyapunov coefficient
$\gamma(E)$, through the asymptotic relation
\begin{equation}
\gamma(E)=[\mathcal{L}_{loc}(E)]^{-1}
=\lim_{n_s\rightarrow\infty}\frac{1}{n_s d}\ln\left|
\frac{G_{n_s,n_s}(E)}{G_{1,n_s}(E)}\right|.
\label{eq-lyap}
\end{equation}
The results shown in the second column of Fig. \ref{fig1} have been computed for
the case $n_s=200$, but we have checked that the values obtained do not change
significantly by increasing the value of $n_s$ up to $1000$.
Analogously to the 
continuous case, we observe that all states are localized. In the limit of weak disorder, the localization length at the center of the spectrum, 
$\mathcal{L}_{loc}(E=0)$, is 
quite large, of the order of 50 lattice sites. By increasing the strength of 
the disorder, $\mathcal{L}_{loc}(E=0)$ decreases 
significantly only for small values of the correlation length ($w=\pi$ and $w=2\pi/3$), and $\mathcal{L}_{loc}(E)$ becomes almost energy-independent 
in the whole band.
Longer-range correlations ($w=2\pi$ and $w=4\pi$) act instead
more efficiently on the edge states.
To better understand these reasults we can refer to the continuous case,
where 
\begin{equation}
\mathcal{L}_{loc}(k)^{-1}\sim \mathcal{L}_{loc}^B(k)^{-1}=\frac{w^2}{8k^2}
\mathcal{S}_{2k},
\label{Born}
\end{equation} 
in the Born approximation 
(see for instance \cite{Kuhl}). 
From Eq. (\ref{Born}) we can expect to observe (i) a decrease of the 
localization length for large values of $w$ 
in the limit $k\rightarrow 0$, since 
$\mathcal{L}_{loc}^B(k\rightarrow 0)\sim k^2/(s^2 w)$, 
and (ii) an increase in the localization length for $k>\kappa/2$, namely where
the Born approximation is no longer valid \cite{Gurevich2009,Luga09}.
Since in the TB case there is a $k\rightarrow \pi/d-k$ symmetry in 
the DOS due to the presence of the underlying lattice and 
correlations act symmetrically with respect to the center of the spectrum,
the observation (ii) leads to the conclusion that, 
if one wants to experimentally 
observe Anderson localization in the whole low-energy band,
 $\kappa/2$ should be greater than $\pi/2d$, thus 
$w$ should be lower than $4$ lattice steps. 
This finding, deduced from propagation in the continuous space, is 
in good agreement with the numerical results for the TB case.

\section{Conclusions}
In this article we have studied the effetcs of a speckle potential on 
the spectrum of a quantum particle (or a non-interacting wave)
in a lattice potential. At fixed, large disorder strength 
($s=10|t|$), the localization efficacy of the speckle potential 
depends strongly on the width of the 
auto-correlation function $w$. Large values of $w$ enhance localization
at very low energies and at the edge of the Brillouin zone.
Shorter-range correlations ($w<4$ lattice sites) act more efficiently
on the center of the spectrum.
More generally, our results show that a speckle superposed to an optical lattice
is a suitable potential to study Anderson localization in the tight-binding 
regime: analogously to the continuous case, speckle correlations deeply modify the behaviour of the localization length as a function of the energy, but do not induce an insulator-metal transition.

\vspace{1cm}
This work was supported by the CNRS and the TUBITAK (exchange of researchers, grant No. 24543). P.V. aknowledges G. Modugno for fruitful discussions, and 
warmly thanks the 
Condensed Matter group of the Trakya University for the kind hospitality. 


\end{document}